\documentstyle[aps,12pt]{revtex}
\begin{document}
\title{A differential geometric approach to singular perturbations}
\author{F. Jamitzky}
\address{Max-Planck-Institut f\"ur extraterrestrische Physik,\\
85740 Garching, Germany\\
Tel. ++49-3299-3386, Fax ++49-89-3299-3569\\
e-mail: f.jamitzky@mpe-garching.mpg.de}
\maketitle

\begin{abstract}
A differential geometric approach to singular perturbation theory is
presented. It is shown that singular perturbation problems such as
multiple-scale and boundary layer problems can be treated more easily on a
differential geometric basis. A general method is proposed based on
differential forms and Lie-derivatives. Examples from multiple scale theory,
boundary layer theory and WKB-theory are given and it is demonstrated that
without the a priori knowledge of the scaling behaviour of the problem the
correct asymptotic expansion can be derived with the aid of differential
forms. The method is well suited for a mechanical implementation in computer
algebra programs.


\end{abstract}

\section{Introduction}

The analysis of boundary layer problems and multiple scale phenomena which
are generalized under the notion of singular perturbation problems has
played a significant role in applied mathematics and theoretical physics
[1,2]. Regular perturbation theory is often not applicable to various
problems due to resonance effects or the cancellation of degrees of freedom.
In order to obtain a uniformly valid asymptotic expansion of a solution for
these singular problems a whole bunch of methods has been developed such as
boundary layer expansions, multiple scale methods, asymptotic matching,
stretched coordinates, averaging and WKB-expansions. Although these methods
work well for the respective cases, a general theory that unifies all these
methods is still lacking [4]. A general applicable method would be highly
desirable in order to obtain a mechanical scheme that can be easily
implemented in algorithmic programming languages.

A physical system often involves multiple temporal or spatial scales on
which characteristics of the system change. In some cases the long time
behaviour of the system can depend on slowly changing time scales which have
to be identified in order to apply multiple scale theory. The choice of the
slowly or fast changing scales is a nontrivial task, which presumes a good
understanding of the physical behaviour of the system and can sometimes only
be justified by the final result. A naive expansion in a power series of the
small parameter is often prevented by the appearance of resonant terms in
higher orders. These terms have to be compensated by the introduction of
counterterms.

Boundary layers are also a common feature of singular perturbed systems. In
these cases higher order derivatives disappear in the unperturbed equations
which leads to the cancellation of degrees of freedom of the system and
finally in small regions where the system changes rapidly. The
identification of the fast changing scales is crucial for the solution of
the perturbation problem and is a subtle task, sometimes involving delicate
arguments.

In this Letter a geometrical approach is presented which allows a mechanical
algorithm for the determination of the rapid or slow scales on which the
system changes. For that purpose the system of ordinary differential
equations is interpreted as a system of differential forms. It is shown that
a small perturbation of the system of differential forms can be viewed as a
deformation of the solution manifold in configuration space. By searching
for a Lie-derivative that transforms the unperturbed differential forms into
the perturbed forms a mapping in configuration space is defined that can be
used to map the solution of the unperturbed system into the an approximate
solution of the perturbed system in order to obtain uniformly valid
asymptotic expansions of the solution and thus determining the secular
scales. The method is exemplified for common problems like multiple scale
problems, boundary layer problems and WKB-expansions.

\section{General Theory}

Assume a system of ordinary differential equations in the $n$-dimensional
configuration space $R^n$ of the following form: 
\begin{equation}
\dot{x}_i=F_i\left( x_1,...,x_n\right)
\end{equation}

where the dot denotes the derivative by the independent parameter $t$. The
system is solved by a $n$-dimensional curve $C:x_i\left( t\right) $
parametrized by $t$. It is easily to shown that all ordinary differential
equations can be written in this form [3,5]. The system of equations can be
written as a system of equations for differential forms: 
\begin{equation}
\omega _{ij}=F_i\left( x_1,...,x_n\right) \;dx_j-F_j\left(
x_1,...,x_n\right) \;dx_i=0
\end{equation}

The differential forms $\omega _{ij}$ establish a basis for a subspace $S$
of the space of all differential forms in $R^n.$ Every solution curve $C$
then has the following property: 
\begin{equation}
\int\limits_Cf\left( x\right) \omega =0
\end{equation}

for all elements $\omega $ of the subspace $S$ and for all scalar functions $%
f$ on $R^n.$ This property can be used as a very general definition of a
solution manifold of a system of differential forms and can be easily
generalized to higher order differential forms and closed ideals of
differential forms.

Now we consider a perturbed subspace $S^{\prime }$ of a subspace $S_0$. Let $%
\omega _0^{\left( i\right) }$ be a basis of an unperturbed subspace $S_0$
and $\omega ^{\left( i\right) }=\omega _0^{\left( i\right) }+\varepsilon
\omega _1^{\left( i\right) }$ be a basis of the perturbed subspace $%
S^{\prime }$ where $\varepsilon $ is a small parameter. The condition for a
solution curve of the perturbed system $S^{\prime }$ can be written as: 
\begin{equation}
\int\limits_Cf\left( x\right) \left( \omega _0+\varepsilon \omega _1\right)
=0
\end{equation}

for all elements $\omega =\omega _0+\varepsilon \omega _1$ of the subspace $%
S^{\prime }$ and for all scalar functions $f$ on $R^n.$ We assume that there
exists a mapping $\Phi $ of $R^n$ into $R^n$ that maps the solution curve $%
C_0$ of the unperturbed system into the solution curve $C=\Phi \left(
C_0\right) $ of the perturbed system. If the mapping fulfills the condition: 
\begin{equation}
\Phi ^{*}\left( \omega _0^{(i)}+\varepsilon \omega _1^{(i)}\right) =\omega
_0^{(i)}+\varepsilon \sum_j\lambda ^{(i,j)}\omega _0^{(j)}  \label{phi-cond}
\end{equation}

with appropriate functions $\lambda ^{(i,j)}\left( x\right) $ then one
obtains: 
\begin{equation}
\int\limits_{\Phi (C_0)}f\;\omega ^{\left( i\right) }=\int\limits_{C_0}\Phi
^{*}\left( f\omega ^{\left( i\right) }\right) =\sum_j\int\limits_{C_0}\left(
\Phi ^{*}\left( f\right) \delta _j^i+\varepsilon \lambda ^{(i,j)}\Phi
^{*}\left( f\right) \right) \omega _0^{(j)}=0
\end{equation}

By expanding the mapping in a power series with the small parameter $%
\varepsilon $ one obtains $\Phi ^{*}=1+\varepsilon \;{\bf L}+{\rm O}\left(
\varepsilon ^2\right) $ where ${\bf L}$ is an appropriate Lie-derivative,
one obtains in first order: 
\begin{equation}
\omega _1^{(i)}+{\bf L}\omega _0^{(i)}=\sum_j\lambda ^{(i,j)}\omega _0^{(j)}
\label{lie-cond}
\end{equation}

This is a linear equation for the Lie-derivative ${\bf L}$ which we can try
to solve. It defines Lie-derivatives that map solutions of the unperturbed
system $S_0$ into solutions of the perturbed system $S^{\prime }$. This
method can be applied also to singular perturbation problems where a
straightforward regular expansion is not available.

\section{Boundary Layer Analysis}

The first example of a singular problem that can be solved with the new
method is given by the following boundary layer problem [1]: 
\begin{equation}
\varepsilon y^{\prime \prime }+y^{\prime }+y=0  \label{bl1}
\end{equation}

with $\varepsilon \ll 1$ and the boundary conditions $y\left( 0\right) =0$
and $y^{\prime }\left( 0\right) =1$ . The prime denotes differentiation by $x
$. This is a common boundary layer problem introducing a layer of thickness $%
\varepsilon .$ The layer thickness is usually obtained from a dominant
balance argument. The system is then rescaled to obtain a inner region
equation which is afterwards matched to the outer region solution. The
choice of the thickness of the boundary layer is justified in the end of the
calculation by the existence of a uniform solution.

In the proposed method such a rescaling is not necessary.. The original
equation is written as a system of differential forms: 
\begin{equation}
dy-z\;dx=0
\end{equation}
\begin{equation}
dy+y\;dx=-\varepsilon \;dz
\end{equation}

The unperturbed differential forms are given by $\omega _0^{(1)}=dy-z\;dx,$ $%
\omega _0^{\left( 2\right) }=dy+y\;dx$ and $\omega _1^{\left( 1\right) }=0,$ 
$\omega _1^{\left( 2\right) }=-dz$ are the perturbation, respectively. The
unperturbed system can be rewritten as $dx=0$ and $dy=0$ which has the
solution $x=x_0$ and $y=y_0.$ This zero-order solution is clearly a singular
solution. In deviates strongly from the zero order solution usually obtained
from the equation $y^{\prime }+y=0$. This difference is crucial in the
present approach. 

After a tedious but straightforward calculation one obtains a Lie-derivative
that maps the unperturbed differential forms into the perturbation: 
\begin{equation}
{\bf L}=\ln \left( y+z\right) \partial _x+\left( z-y\ln \left( y+z\right)
\right) \partial _y
\end{equation}

One easily checks the transformation properties:
\begin{equation}
\left( 1+\varepsilon {\bf L}\right) \left( dy+y\;dx\right)
=dy+y\;dx+\varepsilon \;dz+\varepsilon f_i\omega _i^{\left( 0\right) }
\end{equation}

\begin{equation}
\left( 1+\varepsilon {\bf L}\right) \left( dy-z\;dx\right)
=dy-z\;dx+\varepsilon g_i\omega _i^{\left( 0\right) }
\end{equation}

where $f_i$ and $g_i$ are functions of $x,y$ and $z$ as defined in equation (%
\ref{phi-cond}). From the zero-order solution $x=x_0$ one obtains:
\begin{equation}
\left( 1+\varepsilon {\bf L}\right) \left( x-x_0\right) =x-x_0+\varepsilon
\ln \left( y+z\right) 
\end{equation}

After the substitution $z=y^{\prime }$ the solution can be written as:
\begin{equation}
x=x_0-\varepsilon \ln \left( y+y^{\prime }\right) 
\end{equation}

and thus by substituting and $A=\exp \left( x_0/\varepsilon \right) $ one
obtains the following first-order differential equation: 
\begin{equation}
y^{\prime }+y=Ae^{-x/\varepsilon }  \label{bl2}
\end{equation}

The effect of the mapping from the unperturbed system to the perturbed
system is now clearly demonstrated. In the limit $\varepsilon \rightarrow 0$
the term with the second derivative in equation (\ref{bl1}) vanishes and the
order of the equation changes which implicates a boundary layer which
becomes infinitely thin for small $\varepsilon .$ The transformed system (%
\ref{bl2}) still possesses the additional degree of freedom for small $%
\varepsilon $ and one observes how the singularity forms when $\varepsilon $
becomes small. Equation (\ref{bl2}) can be easily solved in order to obtain

\begin{equation}
y=Ae^{-x}\int e^{x\left( 1-1/\varepsilon \right) }dx+Be^{-x}
\end{equation}

and finally 
\begin{equation}
y=\bar{A}e^{-x/\varepsilon }+Be^{-x}
\end{equation}

which is the correct general uniform first-order approximation. It should be
emphasized that the present approach gives the correct scaling behaviour in
the boundary layer without the need for a rescaling of the independent
variable which is necessary in the usual asymptotic matching procedure or in
the renormalization group approach. Thus the presented method is superior to
the conventional methods in that it chooses selfconsistently the scaling of
the independent and dependent variables in order to obtain a regular
expansion.

\section{Multiple Scales}

Another problem very common in mathematical physics is the appearance of
multiple scales on which the system changes. An example for this behaviour
is given by the following linear oscillator with a nonlinear damping term
[1]: 
\begin{equation}
y^{\prime \prime }+y+\varepsilon y^{\prime 3}=0
\end{equation}

The prime shall denote differentiation by the time variable $t$ . The system
possesses a short time scale of the oscillation and a long time scale of the
damping. The amplitude of the oscillation slowly decays with time and a
frequency shift might be observed.

The unperturbed system is just an ordinary linear oscillator with frequency
unity. The zero-order solution is given by $y=R\cos \left( t+\theta \right) $
with arbitrary constants $R$ and $\theta .$ In order to simplify the
calculations we introduce new dependent variables $R$ and $\theta $ defined
by $y=R\cos \left( t+\theta \right) $ and $y^{\prime }=-R\sin \left(
t+\theta \right) .$ A basis for the perturbed subspace $S^{\prime }$ is
given by: 
\begin{equation}
dR-\varepsilon R^3\left( \frac 38-\frac 12\cos \left( 2\left( t+\theta
\right) \right) +\frac 18\cos \left( 4\left( t+\theta \right) \right)
\right) dt  \label{r-eq}
\end{equation}
\begin{equation}
d\theta +\varepsilon R^2\left( \frac 14\sin \left( 2\left( t+\theta \right)
\right) -\frac 18\sin \left( 4\left( t+\theta \right) \right) \right) dt
\label{th-eq}
\end{equation}

The zero-order forms are just $dR$ and $d\theta $ while the perturbations
are given by the terms with the front factor $\varepsilon $ of equations (%
\ref{r-eq}) and (\ref{th-eq}). The system can further be simplified by
introducing the variable $u=1/R^2$ giving rise to the basis of the perturbed
system: 
\begin{equation}
du-2\varepsilon \left( \frac 38-\frac 12\cos \left( 2\left( t+\theta \right)
\right) +\frac 18\cos \left( 4\left( t+\theta \right) \right) \right) dt
\end{equation}
\begin{equation}
ud\theta +\varepsilon \left( \frac 14\sin \left( 2\left( t+\theta \right)
\right) -\frac 18\sin \left( 4\left( t+\theta \right) \right) \right) dt
\end{equation}

A Lie-derivative that transforms the zero-order differential forms into the
perturbing differential forms can be obtained as: 
\begin{eqnarray}
{\bf L} &=&-\left( \frac 34t-\frac 12\sin \left( 2\left( t+\theta \right)
\right) +\frac 1{16}\sin \left( 4\left( t+\theta \right) \right) \right)
\partial _u+ \\
&&\frac 1u\left( \frac 18\cos \left( 2\left( t+\theta \right) \right) -\frac %
1{32}\cos \left( 4\left( t+\theta \right) \right) \right) \partial _\theta 
\end{eqnarray}

The transformed zero-order solution is then given by: 
\begin{equation}
u=u_0+\varepsilon \left( \frac 34t-\frac 12\sin \left( 2\left( t+\theta
\right) \right) +\frac 1{16}\sin \left( 4\left( t+\theta \right) \right)
\right) 
\end{equation}
\begin{equation}
\theta =\theta _0-\varepsilon \frac 1{u_0}\left( \frac 18\cos \left( 2\left(
t+\theta \right) \right) -\frac 1{32}\cos \left( 4\left( t+\theta \right)
\right) \right) 
\end{equation}

For large $t$ the $\sin $ and $\cos $ terms can be neglected and one obtains
as the final approximation: 
\begin{equation}
R\left( t\right) =\frac{R_0}{\sqrt{1+\frac 34R_0^2\varepsilon t}}
\end{equation}

which is the correct uniform approximation for large times as obtained from
the multiple scale method. The time scale $\varepsilon t$ enters naturally
by the differential forms method and there is no need for an a priori
rescaling of the time variable or the introduction of multiple time scales.

\section{WKB-expansion}

The last example we give is also a very standard problem in mathematical
physics and one of the building blocks of wave mechanics, the
WKB-approximation of linear differential equations. The standard WKB-problem
is given by the following ordinary differential equation [1]: 
\begin{equation}
\varepsilon ^2y^{\prime \prime }+\Omega ^2\left( x\right) y=0
\end{equation}

The basis of differential forms of the unperturbed subspace $S_0$ associated
to this problem and the perturbing forms are given by $\omega
_0^{(1)}=dy-z\;dx,$ $\omega _0^{\left( 2\right) }=\Omega ^2\left( x\right)
y\;dx$ and $\omega _1^{\left( 1\right) }=0,$ $\omega _1^{\left( 2\right) }=dz
$. The basis can be rewritten in order to exhibit the exact nature of the
zero-order forms as: 
\begin{equation}
\omega ^{\left( 1\right) }=\;dx+\varepsilon ^2\frac{dz}{\Omega ^2\left(
x\right) y}
\end{equation}
\begin{equation}
\omega ^{\left( 2\right) }=d\left( y^2\right) +\frac{\varepsilon ^2}{\Omega
^2\left( x\right) }d\left( z^2\right) 
\end{equation}

By introducing the new variables $u=y^2$ and $v=z^2$ a Lie-derivative that
transforms the zero-order basis into the perturbation can be found which
reads as: 
\begin{equation}
{\bf L}=\frac 1{\Omega ^2\left( x\right) }\left( \sqrt{\frac vu}\partial
_x+v\partial _u\right) 
\end{equation}

By using the zero-order solutions $x=0$ and $y=0$ one obtains for the
first-order perturbed equations: 
\begin{equation}
y^{\prime }=\pm \frac i\varepsilon \Omega \left( x\right) y
\end{equation}

and thus recovers the WKB-approximation for the initial problem: 
\begin{equation}
y=e^{\pm \frac i\varepsilon \int \Omega \left( x\right) dx}
\end{equation}

This solution shows again that the differential geometric method simplifies
the task of obtaining a uniform solution for a singular problem.

\section{ Conclusions}

In this Letter we have shown that the differential geometric method is a
superior tool in order to obtain uniformly valid asymptotic expansions of
singular perturbation problems. A basic problem of singular perturbation
theory is the appearance of different scales on which the dependent
variables change. The identification of these scales is a nontrivial process
which requires a deep understanding of the problem under consideration. The
correct choice of the scaling of the independent variables strongly
influences the region of validity of the asymptotic solution. The appearance
of multiple scales can further complicate the task of finding a uniformly
valid approximation of the solution. The method presented in this Letter
facilitates this task and gives the correct expansions in a very natural
way. We have given three classical examples from singular perturbation
theory and have shown how easily uniformly valid approximations can be
obtained with the new method. The method is based on very general
geometrical considerations and thus can be easily extended to more
complicated problems involving partial differential equations (work in
progress).

\end{document}